# MapColorAI: Designing Contextually Relevant Choropleth Map Color Schemes Using a Large Language Model

Nai Yang[a], Yijie Wang[a], Fan Wu[a], and Zhiwei Wei[b,c]*

[a]*School of Geography and Information Engineering, China University of Geosciences, Wuhan, Hubei, China;* [b]*School of Geographic Sciences, Hunan Normal University, Changsha Hunan, China;* [c]*Hunan Key Laboratory of geospatial big data mining and application, Hunan Normal University, Changsha Hunan, China.*

*Corresponding author: 2011301130108@whu.edu.cn

# MapColorAI: Designing Contextually Relevant Choropleth Map Color Schemes Using a Large Language Model

**Abstract**: Choropleth maps, which utilize color schemes to visualize spatial patterns and trends, are simple yet effective tools for geographic data analysis. As such, color scheme design is a critical aspect of choropleth map creation. The traditional coloring methods offered by GIS tools such as ArcGIS and QGIS are not user-friendly for non-professionals. On the one hand, these tools provide numerous color schemes, making it hard to decide which one best matches the theme. On the other hand, it is difficult to fulfill some ambiguous and personalized coloring needs of users, such as requests for 'summer-like' map colors. To address these shortcomings, we develop a novel system that leverages a large language model and map color design principles to generate contextually relevant and user-aligned choropleth map color schemes. The system follows a three-stage process: Data processing, which provides an overview of the data and classifies the data into meaningful classes; Color Concept Design, where the color theme and color mode are conceptualized based on data characteristics and user intentions; and Color Scheme Design, where specific colors are assigned to classes based on generated color theme, color mode, and user requirements. Our system incorporates an interactive interface, providing necessary visualization for choropleth map color design and allowing users to customize and refine color choices flexibly. Through user studies and evaluations, the system demonstrates acceptable usability, accuracy, and flexibility, with users highlighting the tool's efficiency and ease of use.
**Keywords**: Choropleth Map; Color Design; Large Language Model; Data Classification; Cartography.

## 1. Introduction

Maps blend science and art, with artistic elements like color playing a crucial role in enhancing their effectiveness for communication, evoking emotions, and setting the atmosphere (Brewer, 1994; Shive & Francis, 2013). Among various map types, choropleth maps are particularly popular for thematic data visualization, using color gradients to represent data values across geographic regions and effectively highlight spatial patterns (Lei et al., 2023). The success of choropleth maps heavily relies on selecting color schemes that align with both the data and the map's purpose (Brewer et al., 1997; Sun et al., 2015). Therefore, developing tools that help users design tailored choropleth map colors while following cartographic conventions has long been a priority in map design.

To support effective map color design, numerous guidelines have been developed, primarily based on the visual variables first proposed by Bertin (1974) and later refined by scholars like MacEachren (2004), Silver et al. (2013), and Zhou & Hansen (2015). These studies emphasize aligning color choices with data measurement levels: using hue for qualitative data, and value or saturation for ordinal or numerical data (Armstrong et al., 2003; He et al., 2016). Furthermore, the color selection should follow conventional associations (e.g., blue for water, green for vegetation) and adhere to official standards (Wei et al., 2018; Xue et al., 2024), while maintaining sufficient visual contrast for graphical symbols (Chesneau, 2011; Brychtova & Coltekin, 2015). These principles have become essential guidelines for map color design.

Despite these guidelines, users often struggle to create their preferred color schemes. To address this challenge, researchers like Brewer et al. (2003) introduced color templates that offer practical solutions by summarizing widely accepted color combinations. These templates include qualitative schemes with distinct hues for categories and sequential schemes varying in lightness to differentiate data classes. Harrower and Brewer (2003) further advanced this approach by developing ColorBrewer.org, an online tool allowing designers to explore and compare color schemes for thematic maps. Later, Christophe (2011) introduced a knowledge base for map color specifications, integrating insights from visual perception, cognitive science, graphic semiotics, cartography, and art to provide tailored guidelines for color design. Many modern GIS tools, including ArcGIS and QGIS, have also incorporated these templates to assist users in the color design process (Kristian, 2020). Moreover, the advent of deep learning has led to data-driven techniques, such as transferring color schemes from images to maps (Wu et al., 2022) and evaluating the aesthetic quality of map colors (Wu et al., 2024).

However, these automated approaches often fall short of being user-friendly, particularly for non-professionals. On the one hand, users are frequently confronted with a vast array of color schemes, making it challenging to select the one that best matches the map's thematic context. On the other hand, these systems struggle to address more nuanced and personalized color requests, such as a user seeking a 'summer-like' color scheme for a map. This highlights the need for more flexible and interpretable solutions that bridge the gap between automated color generation and user-driven customization. Given that user intentions are frequently conveyed through natural language, the advent of the Large Language Model (LLM) opens up new possibilities for innovative color design. Leveraging their powerful generative capabilities and advanced language understanding (Vázquez, 2024), LLM has the potential to greatly enhance the map color design process by delivering color schemes that are both contextually relevant and closely aligned with user preferences. However, two challenges still need to be solved to apply LLM. First, LLM is also inherently an end-to-end approach, producing homogeneous content for identical inputs, which may not be conducive to controllability issues in the design process (Shi et al., 2023). Second, being a generic model, LLM lacks domain-specific knowledge, such as color scheme types, and color psychology, all of which are crucial aspects of choropleth map color design (Zhang et al., 2024).

To address these challenges, we propose the MapColorAI in this approach, (1) Granting users flexibility in expressing their design intent throughout the process (address challenge I), and (2) Integrating diverse domain-specific knowledge in the design process for choropleth map colors (address challenge II). To achieve this goal, we conducted an in-depth investigation and summary of domain-specific knowledge and design workflows for choropleth map color schemes (*Section* 2). Next, we decompose the choropleth map color design process into a series of sequential steps, applying domain knowledge and LLM at each stage. Additionally, we incorporate multi-level interactive features that allow user intervention at each step to enhance system controllability (*Section* 3). Finally, a user study is implemented to demonstrate the usability of the proposed system (*Section* 4) and discussions are provided in *Section* 5.

## 2. Knowledge of choropleth map color design

In this section, we outline the key components required for designing color schemes for choropleth maps, including classification methods, color theory, and design practices. These elements serve as essential guidelines for developing our map color design system using a LLM.

### 2.1 *Classification method*

(1) Number of classes

Before applying a data classification method, cartographers need to determine the number of classes while making a choropleth map. The human cognitive system, which has limitations on processing distinct chunks of information, suggests that the optimal number of classes should be around 7±2 (Miller, 1956). However, this number can be adjusted based on the complexity of the data. For simple datasets, the number of classes is typically limited to between 3 and 7 (Wei et al., 2018; Slocum et al., 2022), while for more complex datasets, the number can be increased. For example, the ColorBrewer tool supports up to 11 colors for diverging schemes and 9 colors for sequential schemes (Brewer et al., 2003). Based on these guidelines, our framework recommends using between 3 and 11 classes, which is consistent with the approach adopted by Lei et al. (2023). Additionally, the system allows users to adjust the number of classes to suit specific needs or preferences.

(2) Classification method

Numerous classification methods for choropleth maps have been developed for various purposes. We instantiate 6 widely used methods according to Lei et al. (2023) in the proposed system based on the cartographic literature, as follows.

**Equal-Intervals**—is one of the most straightforward classification methods. It divides the entire data range into several intervals of equal width (Evans, 1977). The advantage of this method lies in its simplicity and ease of understanding, as users can easily calculate the boundary values for each interval. However, a significant drawback is that it does not account for the actual distribution of the data. This may result in sparse data points in some intervals, while others may be overly crowded, which could negatively affect the interpretability of the map.

**Quantiles**—is a classification method based on data distribution. It arranges the data in ascending order and then divides it into equal parts, ensuring that each part contains the same number of data points (Evans, 1977). The advantage of this method is that it better reflects the distribution characteristics of the data, maintaining approximately equal numbers of data points in each class, even when the data distribution is highly uneven. However, this method may cause the numeric differences between adjacent classes to become excessively large, which can hinder the visualization of subtle trends or variations in the data.

**Jenks-Caspall**—is a classification method based on natural breaks, which uses an optimization algorithm to determine the best classification boundaries. This method minimizes the variance within each class while maximizing the variance between different classes (Jenks & Caspall, 1971). It is particularly useful when the data is unevenly distributed, as it effectively identifies natural clusters within the data. The strength of the Jenks-Caspall method lies in its optimization capabilities, producing classification results that are more reasonable and data-driven. However, since the method requires multiple iterative calculations, it can be computationally intensive when handling large datasets.

**Fisher-Jenks**—is an improved version of the Jenks-Caspall method, developed by Fisher and Jenks. Similar to the Jenks-Caspall method, it is based on natural breaks but uses a more efficient algorithm to reduce computation time (Jenks & Caspall, 1971). The Fisher-Jenks method employs dynamic programming techniques to determine the optimal classification boundaries, ensuring minimal internal variance within each class and maximizing the variance between classes.

**Max-p**—is a classification method based on cluster analysis, aimed at maximizing the internal consistency of each class (Duque et al., 2012). The method groups data points using clustering algorithms to ensure that the data points within each group are as similar as possible, while points in different groups are as dissimilar as possible. This method is particularly suited for classification tasks that require high consistency and can generate reasonable and well-defined class boundaries.

**Pretty-Breaks**—aims to generate classification boundaries that are easy to understand and remember (Lei et al., 2023). It selects ‘neat’ numbers, such as integers, as classification boundaries, making the results more intuitive and user-friendly. However, this method may not fully reflect the actual distribution characteristics of the data and is best suited for scenarios where high precision is not a critical requirement.

(3) Evaluate the classification quality

Given the variety of data classification methods available, it is important to assess the results of classification to help users select the most appropriate method. To achieve this, we utilize the Goodness of Variance Fit (*GVF*), a numerical indicator that is related to perceptual accuracy and measures the quality of classification results in choropleth map design (Brewer et al., 1997; Lei et al., 2023). The *GVF* is defined as follows:

$$GVF = 100 - \frac{SSW}{SST} \times 100 \qquad (1)$$

Where *SST*(Sum of Squares Total) represents the sum of squared deviations of individual data values from the overall mean, and *SSW*(Sum of Squares Within) is the sum of squared deviations of data values within each class from the class mean, with the sum taken across all classes. A *GVF* value of 9.5 or higher is considered indicative of a ‘satisfactorily accurate classification’ (Declerq, 1995).

## 2.2 ***Color theory***

### 2.2.1 *Color system*

Different applications may require different color systems for maps. For example, the RGB color system is typically used for screen displays, while the CMYK color system is preferred for print publications (He et al., 2016). Additionally, for human color perception, commercial software such as ArcPro and QGIS integrate color systems like HSV, HSB, HSL, and CIELab (Wei et al., 2018). Since our system is web-based and designed for screen display, we have also adopted the RGB color system. Notably, the well-known ColorBrewer tool also provides color schemes with this color system.

### 2.2.2 *Color concept*

(1) Color themes

In many design fields, such as architecture and interior design, color themes play a crucial role in defining the overall style (Hou et al., 2024). For instance, an 'elegant' color theme often incorporates muted or soft tones to foster harmony and balance. Similarly, color themes are also integral to map design. He et al. (2016) have outlined commonly used color themes for maps, including 'light', 'moderate', 'strong contrast', 'elegant', and 'neutral'. Building on this, we apply these themes to choropleth map color design, with the potential for further expansion in future applications.

(2) Color moods

Psychological effects and emotional responses may be evoked by different colors. A thoughtful application of color psychology can enhance the expressive impact of maps, guiding user attention and emotions. From a psychological perspective, the influence of color on choropleth map design is primarily reflected in three aspects also known as color moods.

**Temperature Perception**. Colors evoke a sense of warmth or coolness, which can influence how map areas are interpreted. Warm colors, such as red, orange, and yellow, typically evoke feelings of warmth, energy, and forward movement, making them ideal for representing regions with high temperatures, high densities, or greater significance. In contrast, cool colors—such as blue, green, and purple—convey calmness, retreat, and stability, making them more appropriate for depicting areas with lower temperatures, low densities, or background regions.

**Spatial Perception**. Colors can influence the observer's subjective judgment of spatial dimensions, including area size and distance. Colors with higher brightness and saturation tend to create a sense of forward movement, making regions appear larger and more prominent. In contrast, colors with lower brightness and saturation tend to recede, giving the impression that the areas are smaller and farther away. This spatial perception effect can be strategically employed in choropleth map design to establish a visual hierarchy, highlight key areas, and de-emphasize less significant regions, thereby effectively guiding the viewer's attention to the most relevant parts of the map.

**Weight Perception**. In choropleth maps, darker colors are often perceived as heavier and more stable, while lighter colors convey a sense of lightness and freshness. This weight perception is crucial for illustrating hierarchical differences in data, helping to establish visual prominence and subordination. By using this effect, cartographers can enhance the clarity of data classification, ensuring that higher-priority regions stand out, while less significant data is visually subdued.

#### 2.2.3 *Quantitative representation of color*

The choropleth map uses color to represent the data-associated areas. A foundational principle in designing color schemes for choropleth maps is that people tend to associate higher values with darker colors. It follows that if the choropleth needs to show logical ordering of value then rainbow colors must be avoided (Gołbiowska & Çöltekin, 2020). To better reflect the data, two kinds of data or kind of color schemes have been provided by ColorBrewer and have been designed principles in choropleth design (Brewer et al., 2003). For sequential data, smooth transitions from light to dark shades effectively represent low-to-high values (sequential scheme). In diverging data, balanced midpoints and contrasting extremes ensure clarity (diverging scheme). A well-chosen color scheme highlights subtle variations while maintaining overall visual coherence. In the proposed system, we also adopted the two-color scheme (sequential vs diverging) in our system.

### 2.3 *Design practice*

Based on previous research (Slocum et al., 2022; Lei et al., 2023) and interviews with two experts, the workflow for color design in choropleth maps can be divided into three key steps. The system is designed following this established design practice.

(1) **Data processing**. In designing a choropleth map color scheme, the mapmaker first gains an overview of the data, including its content and range. The next step involves selecting an appropriate classification method to group the data into discrete intervals or classes, as well as determining the color scheme type (sequential or diverging) based on the data characteristics. The choice of classification methods— such as equal intervals, quantiles, or natural breaks—can significantly influence the map's visual representation and interpretation. This decision depends on the data distribution and the intended focus, whether it's to highlight trends, outliers, or variations. For instance, when visualizing population density, using natural breaks may be preferable to emphasize meaningful differences between regions, ensuring the map highlights relevant variations effectively.

(2) **Color concept design**. In the second step, designers conceptualize the color concepts including color themes and color moods based on the data and specific user requirements. First, designers associate specific color moods with user requirements, keeping in mind color psychology. Second, designers establish the overall color theme, such as 'light' or 'elegant', providing a cohesive style for the map that aligns with the intended message and user preferences. For instance, a vague user request, such as 'a lively atmosphere for the map', may be interpreted as favoring a warm color mood and a 'strong contrast' color theme, which are known to evoke energy and vibrancy.

(3) **Color scheme design**. The final step involves designing the specific color scheme to be applied to the map. At this stage, the designer selects appropriate colors for each class defined in the classification step. Based on the concept developed in Step 2, the color scheme is refined to reflect the color themes and color moods. For example, if a warm color mood was selected in the concept design phase, colors such as red or orange may be incorporated to visually represent higher values. This stage ensures that the map's color scheme aligns with both the data content and the overall design concept, enhancing the map's clarity and interpretability.

## 3. Methodology

### 3.1 *System overview*

This study aims to investigate the use of LLM-based assistance in designing color schemes for choropleth maps, focusing on transforming cartographers' vague mapping intentions into color schemes while adhering to the fundamental principles of map color design and providing language interaction. Based on the design practices outlined in *Section* 2.3, we developed the 'MapColorAI' system, which supports choropleth map color scheme creation through three stages: data processing (*Section* 3.2), color concept design (*Section* 3.3), and color scheme design (*Section* 3.4), as illustrated in Figure 1. However, unlike traditional workflows, our approach incorporates LLM functionalities into each stage to streamline and enhance the design process. Each stage integrates

LLM capabilities with user interactions (*Section* 3.5), enabling the creation of personalized color schemes while adhering to the specific design principles of choropleth maps.

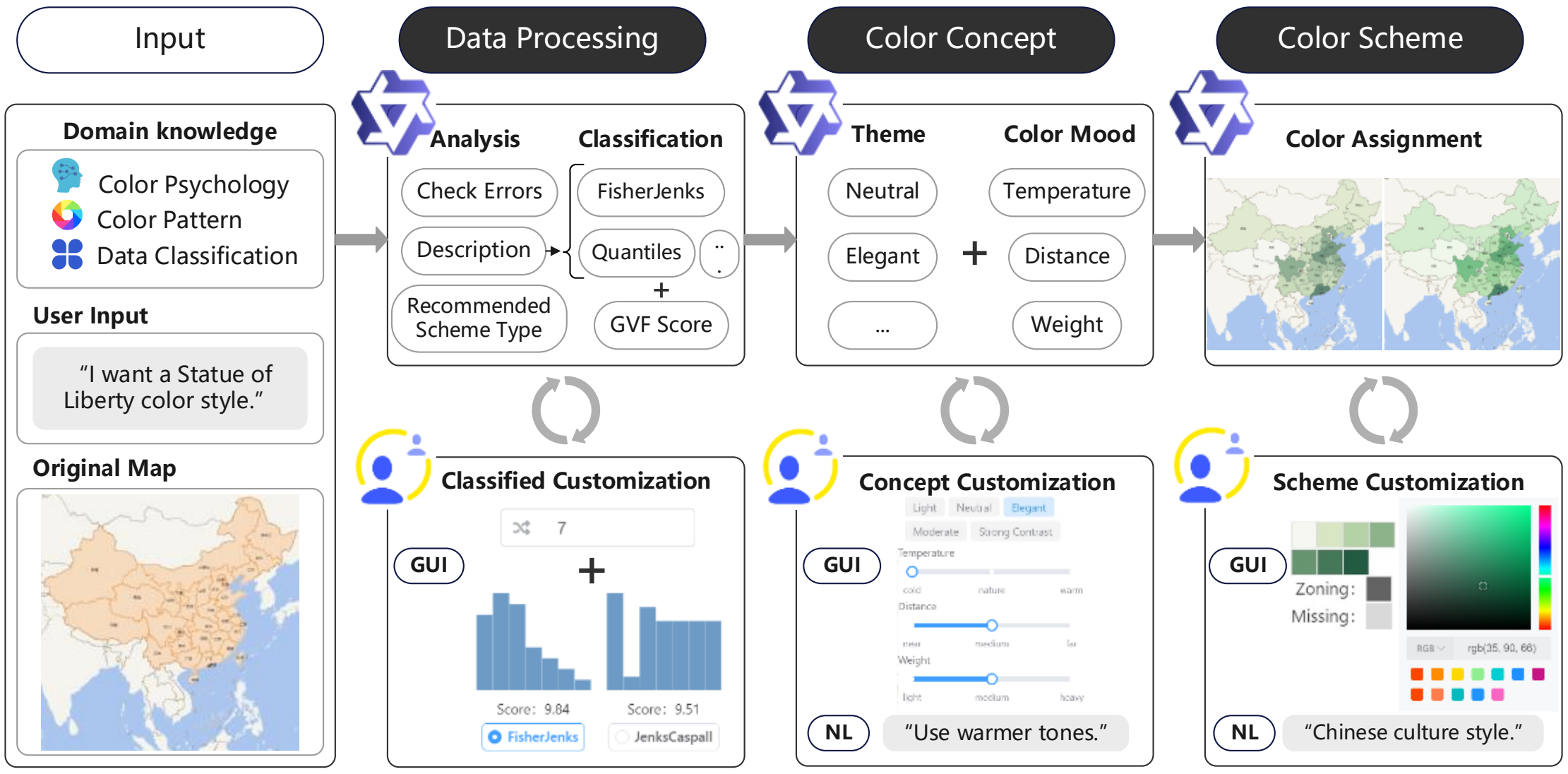

Figure 1. MapColorAI encompasses three stages: (1) Data processing: Data comprehension, data classification, and determining color scheme type; (2) Color concept design: Customization of color themes and color moods; (3) Color scheme design: Generation of specific color schemes and real-time preview.

## 3.2 *Data processing*

In this stage, two sub-processes are implemented, as discussed in *Section* 2.3: data analysis and data classification. The data analysis process identifies potential data errors, provides an overview of the data, and suggests the appropriate color scheme type (sequential vs. diverging) based on the data's characteristics. This is achieved through the use of a large language model (LLM). The data classification process applies commonly used classification methods to categorize the data and recommends the highest-rated method based on an evaluation of the classification result's quality.

### 3.2.1 *Data analysis*

Recent research has demonstrated that LLM is highly effective in understanding statistical data (Tian et al., 2024). Thus, we directly specify the tasks in the prompt for the LLM, without the need for special design to implement the data analysis. The primary objective of this process is to provide the user with an overview of the data and to extract key information that will inform subsequent color design.

The tasks include: (1) Check possible data errors in the upload data, such as missing data or abnormal values. (2) Provide as detailed a description as possible based on the uploaded data, including topics, range, acquisition time, etc., as much as possible. This information such as data topic may be necessary for later color design, for example, ‘blue’ tone may be preferred for a data topic related to ‘water’. The general data description is also an input for later color design. (3) Suggest color scheme type (sequential vs diverging) based on data characteristics. As is not specifically designed for choropleth map creation, we integrate domain knowledge to guide this task. Domain knowledge is set according to He et al. (2016) and Peterson (2020) as follows:

*‘Sequential scheme is ideal for visualizing data with a clear order or magnitude, such as population density or income levels. Diverging scheme is ideal for visualizing data that deviates in two opposite directions from a meaningful midpoint, such as temperature anomalies or percentage change. You can determine the scheme type based on—Does the ranking have a ‘center’ or ‘middle’? If it does, a diverging scheme is appropriate; if not, a sequential scheme is preferred.’*

Here, we take the ‘2023 GDP of Chinese Provinces’ (excerpt) as an example. The data format is JSON, with the names of each region and their corresponding GDP values within the same array item. The data analysis module may output the following response for the above three tasks, as shown in Figure 2. For Task 1: LLM found no errors in the data; for Task 2: The LLM summarized information such as data coverage, maximum and minimum values, and significant differences between regions; for Task 3: It recommends using the ‘Sequential’ color scheme type. As demonstrated in this example, the module provides a comprehensive overview of the data, allowing users to obtain key insights without needing to manually inspect the JSON file.

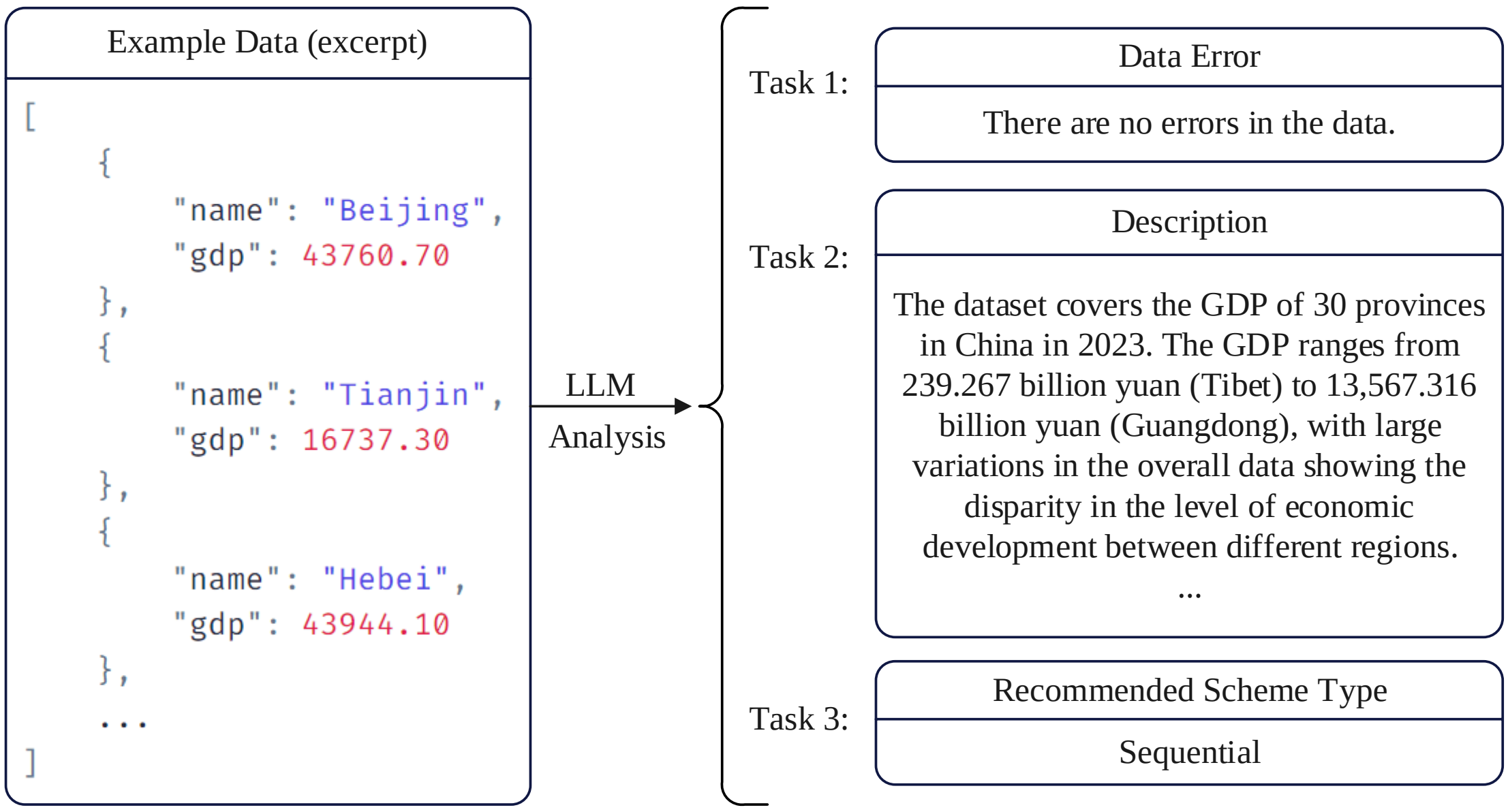

Figure 2. The left side illustrates an example of input data, while the right side presents the results of the LLM completing three data analysis tasks.

3.2.2 *Data classification*

After obtaining an overview of the data, the user proceeds to the data classification step. This is a semi-automatic process in which the LLM is not involved. The user begins by selecting the number of classes, which is restricted to a range between 3 and 11, as determined by the analysis in *Section* 2.1. Once the classification number is chosen, the system applies six commonly used classification methods (Equal-Intervals, Quantiles, Jenks-Caspall, Fisher-Jenks, Max-P, and Pretty-Breaks, as described in *Section* 2.1) to categorize the data. The Goodness of Variance Fit (GVF), defined by Equation 1, is then used to evaluate the quality of the classification (Lei et al., 2023). This score helps the user identify their preferred classification method, with the method achieving the highest score being automatically selected. For instance, the Fisher-Jenks method, as shown in Figure 3, may be selected based on its superior score. To further assist users, the classification results are visualized using histograms, providing a more intuitive understanding of the classification and allowing users to select the method that best meets their specific mapping needs.

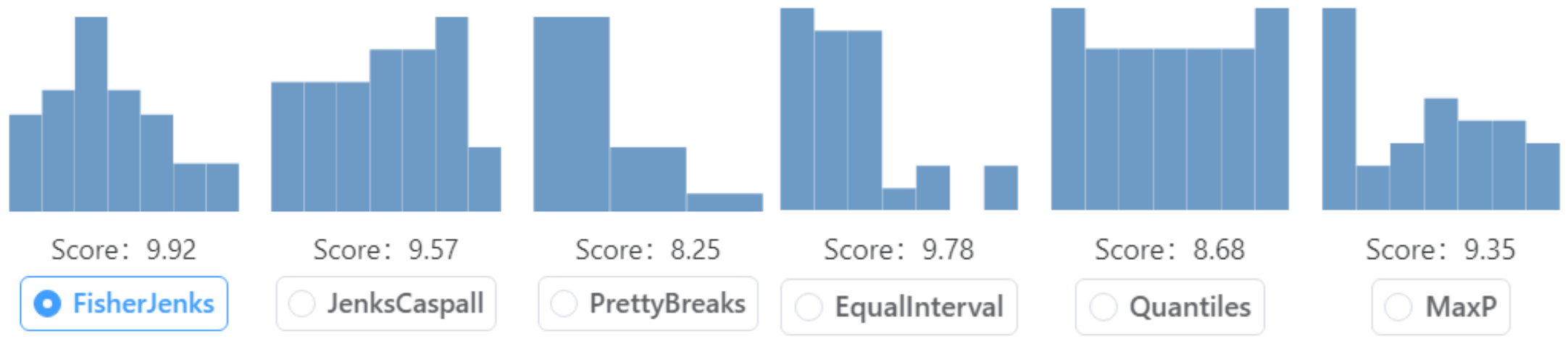

Figure 3. Visualization of the data classification results, illustrating the distribution and selection of classification methods.

## 3.3 *Color concept design*

3.3.1 *Color concept generation*

This step involves translating the user's vague mapping intentions into color concepts, including specific color themes and corresponding color moods. This is achieved by using the LLM with a structured prompt template to guide the generation translation process. The prompt follows the common practice which consists of five parts (Caelen & Blete, 2024), as illustrated in Figure 4.

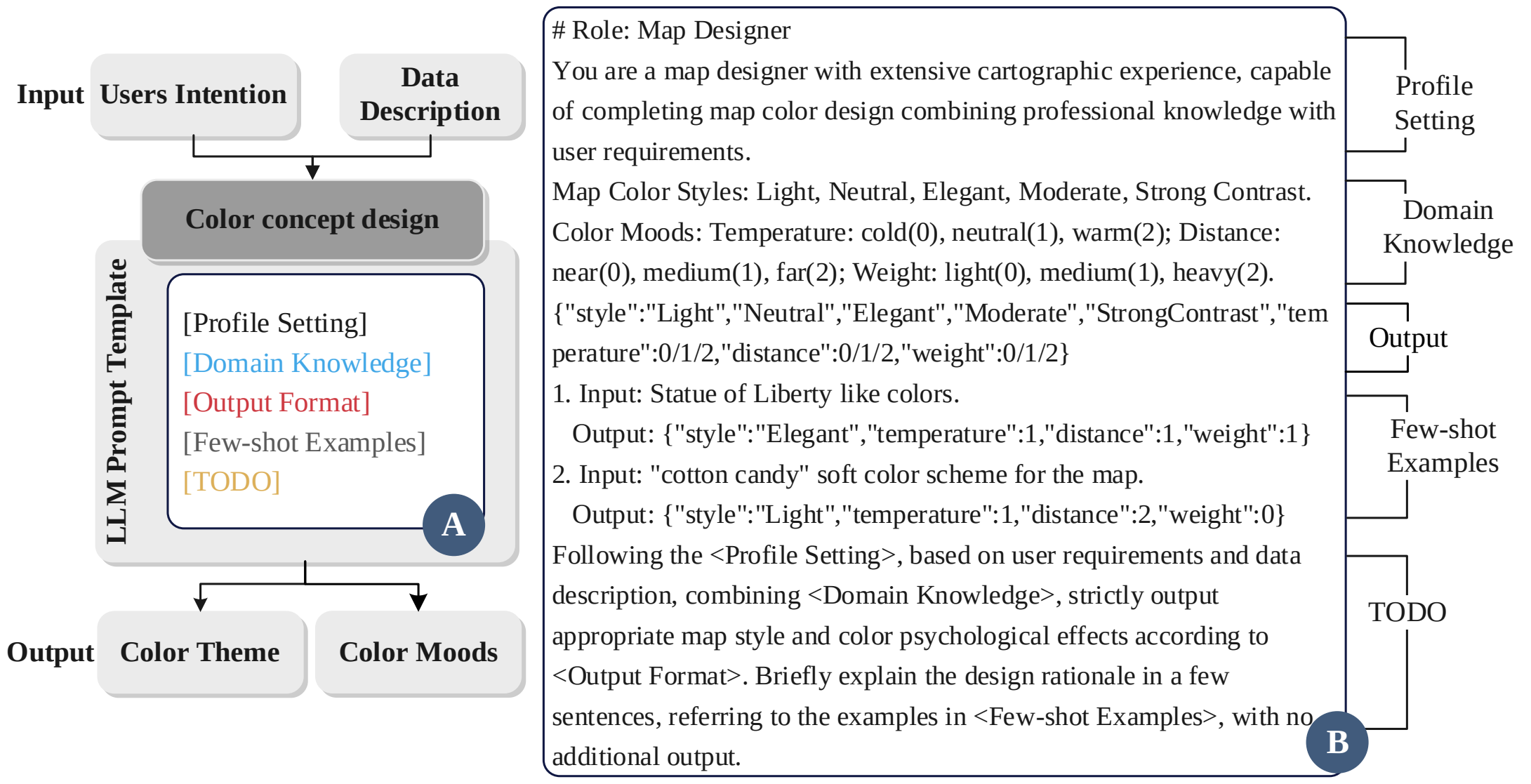

Figure 4. Color concept generation. (A) The prompt template; (B) A specific prompt that integrates data descriptions to transform user vague input into color theme and color moods, taking into account domain knowledge of map color design.

- Data Input: As shown in Figure 4(A), the input for this stage includes the user's vague intent and the data description information generated in the previous stage, enabling the LLM to output content that meets the user's requirements based on the data characteristics.
- Profile Setting: It has been empirically observed that assigning a specific role to an LLM significantly enhances its performance (Caelen & Blete, 2024). As we aim to design a color scheme for the choropleth map, we assign the LLM with a role as map designer, briefly outlining its character's background and skills, to ensure the activation of the model's role-playing capabilities.
- Domain Knowledge: The color concept design, driven by user intentions, is typically carried out by professionals with domain-specific expertise. To ensure that the output aligns with design requirements, we summarize the commonly considered color themes and color moods in choropleth map design, as analyzed in *Section* 2.2.2. The color themes include 'strong contrast', 'light', 'moderate', 'elegant', and 'neutral', which can also be extended in practice. The color moods include 'Temperature Perception', 'Spatial Perception', and 'Weight Perception'. Each attribute is quantified into three levels. To ensure accuracy and consistency, we align these levels' natural language descriptions with specific numerical values, such as 'cold (0), neutral (1), warm (2)' for temperature, 'near (0), medium (1), far (2)' for distance, and 'light (0), medium (1), heavy (2)' for weight.

- Output Format: It is a constraint on the output format of LLM, which ensures a standardized output to facilitate subsequent parsing and utilization.
- Few-shot Example: Research has found that LLM is capable of generating higher-quality results when provided with a small number of sample examples (Liu et al., 2023). In map color design, the users' intentions and their related themes or color moods are usually various and vague, which are difficult to summarize into high-level design rules, such as mapping from the 'Statue of Liberty color style' to an 'elegant' map style. Therefore, these mappings cannot be included in the domain knowledge section. To address this issue, we supplement the domain knowledge by providing users with fuzzy inputs paired with corresponding design concepts, as shown in Figure 4(B). An example of this includes user intentions, map themes, and three key attributes that reflect the desired color moods.
- TODO: In this section, a comprehensive framework for the behavior of the LLM has been established, ensuring that it follows the outlined procedures to accomplish tasks based on the aforementioned content. Additionally, the LLM is required to provide a design rationale for its outputs. The rationale serves to explain the reasoning behind the model's design choices, ensuring that the generated solutions are not only contextually relevant but also align with established principles of map design, user intentions, and color theory. This transparency helps users understand the underlying logic of the LLM's decisions, fostering trust and improving the overall user experience.

As shown in Figure 4(B), an example-generated prompt is illustrated. Here the user wants a 'Statue of Liberty like' map color design, and the LLM uses the prompt transfer it into design concepts, including themes such as 'elegant', and color mood attributes including 'neutral tones', 'medium distance', and 'medium weight'.

3.3.2 *Color concept design customization*

The system supports interactive modifications for color concept design through both graphical interfaces and natural language. On the one hand, map themes are presented in the form of tags, while color mood attributes (temperature, distance, and weight) are displayed via sliders, as shown in Figure 6($B_3$). Users can select other theme tags or modify the intensity of color moods through the interface directly.

Additionally, we also provide a natural language-based interaction method for users with more ambiguous needs, enabling them to refine input information or make adjustments. For example, the 'classic' color theme can be further interpreted as using 'light weight' or 'heavy weight' color schemes and will result in two distinct choropleth maps. Users can clarify their preference by specifying 'classic soft tones', indicating a desire for a more gentle color combination that conveys a sense of stability and tradition, thus modifying the color mode into 'light weight'.

### 3.4 *Color scheme design*

#### 3.4.1 *Color scheme generation*

During this phase, the classified results and color concepts derived from the previous two steps are transformed into specific color schemes. In traditional map color design, this stage is typically domain-specific, where the map maker, drawing on experience or referencing existing color schemes or websites, designs the final color scheme (He et al., 2016; Xi et al., 2023). To address this, two sub-processes are implemented in this stage. First, a color scheme is generated using the large language model (LLM), leveraging the data and color concepts from the first two stages (**color scheme generation via LLM**). Second, to ensure the output color scheme aligns with professional standards and optimally utilizes existing color scheme resources, we provide users with matching color schemes from a database of well-established options (**color scheme matching**).

(1) Color scheme generation via LLM

This process is implemented using the large language model (LLM) through a specialized prompt template, similar to the one used in the color concept design stage. As a result, we provide only a concise description of the prompt in this stage, omitting detailed explanations. The prompt includes domain knowledge, output format specifications, few-shot examples, and behavioral settings, as illustrated in Figure 5(A). Since the dialogue at this stage builds upon the content from Section 3.2 (Color Concept Design), role settings are also incorporated within the prompt. The generation process capitalizes on the LLM's ability to map semantically similar words, a feature shaped by both the model's pre-training and the specific context of the prompt. This enables the LLM to effectively align color concepts and data topics with established knowledge in the field of map color design, ensuring both semantic and thematic consistency. For instance, as shown in Figure 5($B_1$), based on the 'warm' color mood in the color concept, the LLM proposed a color scheme dominated by red and yellow tones, avoiding cooler shades like blue or green. This selection aligns with the ‘Sequential’ color scheme type, transitioning from light yellow to deep red to represent different data levels through varying shades of a similar hue. Additionally, the ‘Strong contrast’ color theme is evident in the results, which accentuates the differences between data levels, making them more visually distinct.

(2) Color scheme matching

The ColorBrewer color schemes are widely recognized for their effectiveness in choropleth map design (Lei et al., 2023). To enhance the professionalism of the output color schemes, we incorporate ColorBrewer’s color schemes as a reference database. A total of 207 ColorBrewer color schemes, downloaded from the ColorBrewer website, are used as the base for comparison. The system calculates the difference between the output color scheme generated by the LLM and each scheme from the ColorBrewer database, with the most similar color scheme being recommended to the user, as shown in Figure 5. To compare the color schemes, we first filter out those with the same number of colors as the output scheme, ensuring a more accurate match. As both the LLM-generated and ColorBrewer schemes are represented in RGB format (which is commonly used for screen displays but not aligned with human color perception), we convert both sets of colors from RGB to the CIELab color space—a color system designed to reflect human color perception (Harrower & Brewer, 2003). The comparison is then conducted by calculating the color differences between the two schemes in the CIELab color space. The difference ( $\Delta E$ ) between two colors is quantified as the Euclidean distance between their respective points in the CIELab color space, and defined as follows.

$$\Delta E = \sqrt{(L_1 - L_2)^2 + (a_1 - a_2)^2 + (b_1 - b_2)^2} \quad (2)$$

Where $(L_1, a_1, b_1)$ and $(L_2, a_2, b_2)$ are the two colors in the CIELab color system.

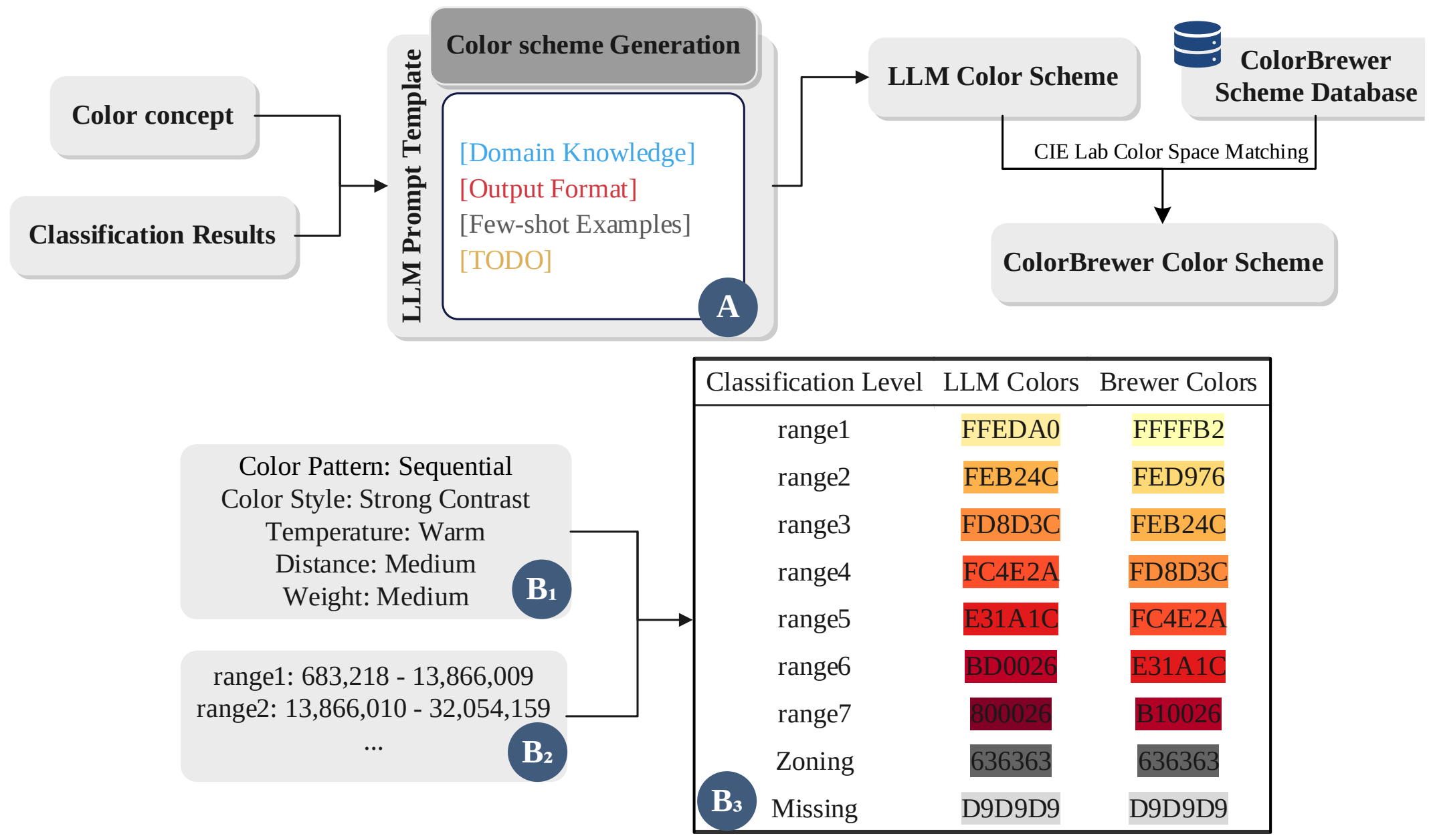

Figure 5. Color scheme generation. (A) The prompt template used for color scheme generation; ($B_1$ and $B_2$) an example of input data; ($B_3$) The LLM-generated color scheme and its most similar ColorBrewer color scheme.

#### 3.4.2 *Scheme Customization*

After generating a color scheme, we provide an interactive customization interface for users, as illustrated in Figure 6($B_4$). This graphical interface enables users to adjust each color individually, with real-time updates that allow for immediate visual feedback. For example, if a user finds it difficult to distinguish between two colors, they can fine-tune attributes such as brightness or saturation to create clearer differentiation. To further simplify the customization process, especially for non-professional designers, we also incorporate natural language interaction, which is similar to the color concept design stage. Users can adjust the color scheme with simple commands. For instance, if a user feels that the overall color scheme is too monotonous or lacks vibrancy, they can simply request, ‘Make these colors more vivid’, and the system will interpret this intent by enhancing the color saturation or introducing contrasting hues to increase visual impact.

### 3.5 *Implement details and interaction design*

#### 3.5.1 *Implement details*

The system backend is built using Flask integrated with Qwen 2.5 API. The architecture is modular, with distinct functions for each LLM operation. These functions execute sequentially, with each stage’s output becoming the next’s input. For each stage, we use few-shot examples that cover map color design styles provided by our collaborating professionals. All the examples and outputs are in a structured format, which makes it easy for the LLM to interpret. Following conventions in few-shot prompting, we provide 2 examples in the color concept design stage and color scheme design stage. These examples strike a balance between providing sufficient context and maintaining brevity for efficient model performance. The data analysis phase uses the Qwen-long model with a single output limit of 6,000 tokens and a context length of 10,000,000 tokens, which facilitates the processing of larger data files. For the other phases, the Qwen-plus model is used, with a maximum single output of 8192 tokens and a context length of 131,072 tokens. Temperature is set to the default value of 1.0 for all stages. The interface is a web-based application crafted using Vue.js, the map is rendered using leaflet, and some other components use the Element Plus component library.

#### 3.5.2 *Interaction design*

We have developed an interactive mapping system that integrates LLM into every step of the mapping process, as illustrated in Figure 6. The interface primarily consists of three views:

(1) **Conversation View** (Figure 6(A)). This view serves as the primary interface for users to interact with the system using natural language. Users can guide the color design process by inputting instructions, such as ‘I want a Statue of Liberty like map’ for generating the color scheme directly, or ‘make the colors brighter’ or ‘increase the contrast’ to refine the color concepts or color schemes. The results are then previewed in the right-hand view. To handle complex tasks, this view supports multi-turn dialogues, allowing users to gradually refine their requirements and receive coherent assistance.

(2) **Color design view** (Figure 6(B)). This view focuses on the interactive modification of various attributes during the color design process. As mentioned earlier, we divide the choropleth map color creation process into three stages. In the first data processing stage, users can upload their mapping data (Figure 6($B_1$)), classify the data, and we provide visual results of six classification methods for users to choose from, automatically recommending the result with the highest GVF score (Figure 6($B_2$)). In the second stage of color creative design, users can select the desired color scheme types and color themes through tags and adjust color psychological effects such as temperature, weight, and distance (Figure 6($B_3$)). In the third stage of color schemes, specific colors generated by LLM in hierarchical order are displayed, along with the closest ColorBrewer color scheme, allowing users to switch between these two schemes for viewing or adjust the specific color using a color modified plates (Figure 6($B_4$)).

(3) **Map view** (Figure 6(C)). This view displays the resulting map using the generated color scheme, allowing users to zoom in, zoom out, and pan the map to inspect the coloring details of different areas. The hierarchical division of the mapping area and the corresponding colors are placed in the legend at the bottom right corner for user reference. The map view is tightly integrated with the other two views, and any color changes are instantly reflected in the map view, facilitating user adjustments and refinements.

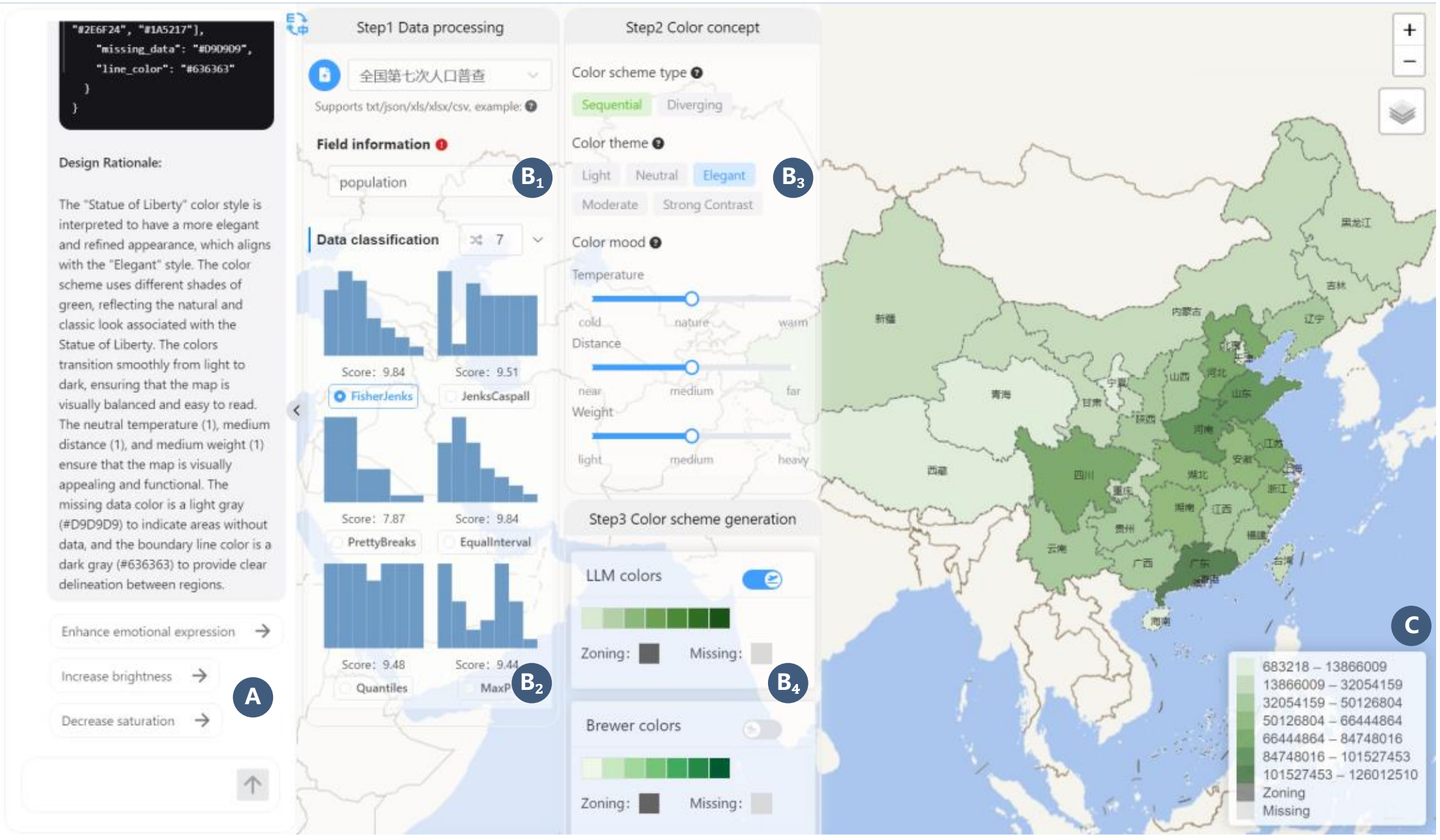

Figure 6. System Interface (an example of ‘Statue of Liberty’ map color design). (A) Conversation View: Users can input their initial design intents and customizations in natural language. (B) Color Design View: Users can fine-tune the intermediate results generated by the LLM. (C) Map View: Users can evaluate the results with reasoning and refine the color assignments based on their preferences.

## 4. User study

To assess the usability of the proposed system, we conducted a user study to assess the system’s usability and satisfaction, including outcome relevance (Q1 and Q2), tool completeness (Q3), ease of use (Q4-Q6), system flexibility (Q7), and overall user satisfaction (Q8-Q10), with the tasks illustrated in Figure 7. Participants were instructed to use the system to generate color schemes for a given dataset. Feedback was then collected through a combination of structured questionnaires and semi-structured interviews.

### 4.1 *Participants*

We conducted a user study with 60 participants (P1-P60) aged between 18 and 44 years. Participants self-identified their gender, with 32 identifying as male and 28 as female. Given that professional knowledge or background in designing color schemes for maps could significantly influence feedback, we assessed participants' familiarity with this subject. They responded to a Likert-scale question: '*Please indicate your level of knowledge [1, 2, 3, 4] regarding color scheme design for a map*.' Responses were categorized into two groups: unfamiliar (levels 1 and 2) and familiar (levels 3 and 4). The results showed an equal distribution, with 30 participants classified as familiar and 30 as unfamiliar.

### 4.2 *Material*

Participants will uniformly use the administrative division map of China as the base map. Four sets of data are prepared for users to create maps, and users can freely choose to create at least one map, with a brief introduction to the data as follows:

- 'Per Capita Disposable Income by Region in 2022 of China', unit (yuan), data field 'income';
- 'GDP of Provinces in 2023 of China', unit (billion yuan), data field 'gdp';
- 'Seventh National Population Census in China', unit (person), data field 'population'.
- 'Total Electricity Generation by Region in 2022 of China', unit (100 million kWh), data field 'generation'.

### 4.3 *Procedure*

The procedure contains three steps, as follows: (1) a tutorial session to make users familiarized with the proposed system, (2) a creation session to experience the authoring process, and (3) a post-study evaluation to collect their feedback on the utility of the system.

**(1) Tutorial**. Participants were introduced to the proposed system through a 15-minute presentation explaining the motivation behind the work. This was followed by a 10-minute demonstration using slides, showcasing the system's key functionalities, such as inputting user requirements, setting the number of classifications, selecting a color mood, designing a color scheme, and interacting with related features. Participants were then encouraged to freely explore the system's functions and interactions, asking questions as needed. Once participants confirmed their familiarity with the tool, we introduced the formal dataset for the user study.

(2) **Creation**. After completing the tutorial, participants were instructed to use the system to design their own map by creating a custom color scheme. They could refer to the tutorial slides and request additional guidance if necessary. Upon completing their designs, each participant shared and explained their map. This creation phase lasted approximately 5 to 10 minutes.

(3) **Post-study Survey and Interview**. Following the creation phase, participants completed a post-study questionnaire featuring 10 questions (Figure 7) using a 5-point Likert scale (1 = strongly disagree, 5 = strongly agree). The 10 questions were used to assess the system's usability and satisfaction, including outcome relevance (Q1 and Q2), tool completeness (Q3), ease of use (Q4-Q6), system flexibility (Q7), and overall user satisfaction (Q8-Q10). Additionally, participants provided demographic information, including age, gender, and familiarity levels, before responding to the system evaluation questionnaire. Finally, we conducted semi-structured interviews to collect qualitative feedback from each participant, offering deeper insights into their experiences and perspectives.

### 4.4 *Result analysis*

As we use the Likert scale to assess the system's usability and satisfaction, the reliability and validity of the survey were first assessed using Cronbach's $\alpha$ coefficient, which was 0.888 (standardized $\alpha = 0.887$), indicating good internal consistency, and a KMO value of 0.894 along with Bartlett's test of sphericity ( $\chi^2 = 251.341$, $p < 0.001$), confirming significant correlations among variables and the suitability of the data for factor analysis. The results indicate the reliability and validity of the survey and the detailed results of different tasks are analyzed as follows.

#### 4.4.1 The overall results

The average scores, standard deviations (SD), and response distributions for each question are presented in Figure 7. As shown, there was a strong level of agreement among participants, with the overall scores indicating high usability of the proposed system. The mean score for outcome relevance (Q1 and Q2) was 4.00 (SD: 0.64) and 4.55 (SD: 0.50), respectively, reflecting that participants found the tool highly relevant to their needs. For tool completeness (Q3), the average score was 4.23 (SD: 0.55), suggesting that users perceived the tool as comprehensive and capable of meeting their requirements. As P2 remarked, 'The system is relatively convenient to use and the tools can meet most color matching needs.'

When evaluating ease of use (Q4-Q6), the average scores ranged from 4.23 to 4.38, with standard deviations between 0.55 and 0.61, indicating that participants found the tool intuitive and user-friendly. Regarding the interface, P22, P25, and P53 specifically commented, 'The interface is concise and aesthetically pleasing.' P3 noted, 'The interface is straightforward, making it easy for me to start using it without a steep learning curve.' For system flexibility (Q7), the average score was 4.13 (SD: 0.57), suggesting that users appreciated the tool's adaptability. As P24 observed, 'The tool can be customized to suit my specific needs, which is an important feature for non-experts.'

Finally, overall user satisfaction (Q8-Q10) yielded average scores of 4.27 (SD: 0.69), 4.33 (SD: 0.60), and 4.32 (SD: 0.65), reflecting a high level of user contentment. P41 remarked, 'Using large language models for color matching is both interesting and convenient, saving a lot of time and effort.'

In summary, the average score for all questions exceeded 4.0, with participants highlighting the tool's accuracy, ease of use, and flexibility as key factors contributing to their positive experience.

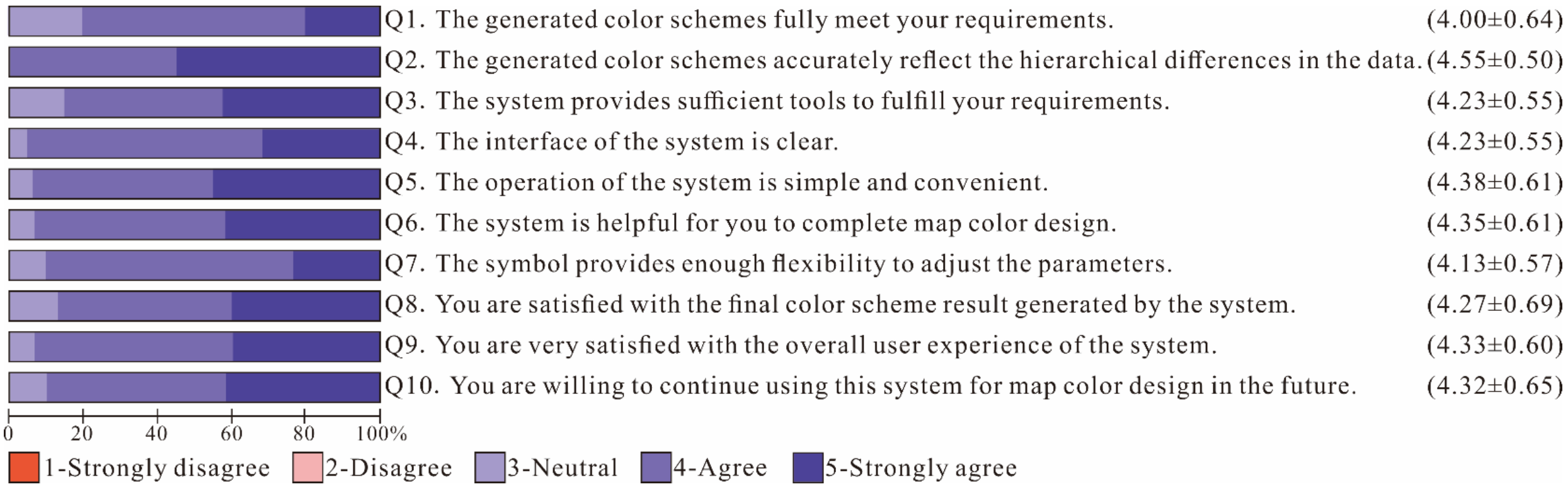

Figure 7. Ratings for system usability and satisfaction on a 5-point Likert scale. The middle column shows the detailed questions. The right column displays the average and standard deviations.

4.4.2 The influence analysis of familiarity (familiar vs unfamiliar)

The statistical analysis was conducted to compare the performance between two groups—Familiar and Unfamiliar—across 10 questions (Q1-Q10). The Shapiro-Wilk test was first employed to assess the normality of the data. Although the *p*-values for all 10 questions were below 0.05, indicating a departure from normality, further analysis based on skewness and kurtosis revealed that each question has a bell-shaped distribution with a peak in the middle and lower ends. These suggest the data could be considered approximately normal. Consequently, a test for homogeneity of variance was performed to assess the equality of variances between the two groups. In cases (Q1-Q3, Q5, Q6, and Q8-Q10) where the homogeneity assumption was met, an independent samples t-test was used to determine if there were significant differences between the groups. In cases (Q4 and Q7) where the homogeneity assumption was violated, Welch's t-test, which does not assume equal variances, was used. The results are shown in Table 1.

The results show that the Unfamiliar and Familiar groups exhibited significant differences in performance in Q1 and Q9, with *p*-values less than 0.05 and Cohen's d values greater than 0.5. These results suggest that the differences, while statistically significant, have moderate practical significance. Such differences may be attributed to the contrasting experiences of the two groups. For example, professionals familiar with pre-set templates may find it challenging to generate color schemes that cater to personalized needs, whereas non-experts may not perceive this issue. In contrast, the new system addresses these challenges effectively. As noted by Participant P5: 'This tool can effectively address the lack of color knowledge among beginners in map design. Many beginners tend to use random color schemes from tools like ArcGIS or MapGIS, and even when there are customization options, adjusting them can be difficult. However, this tool integrates large language models for custom color generation, solving the problem of inaccurate map color choices for non-professionals and demonstrating excellent application effectiveness.' This feedback highlights that experts recognize the potential of the proposed system, particularly its practical value for future use.

Table 1. The influence analysis of familiarity.

| Questions | Factor | Average score | Homogeneity test (*p*-value) | t-Test (*p*-value) | Welch's t-Test (*p*-value) | Cohen's d value |
|---|---|---|---|---|---|---|
| Q1 | Familiar<br>Unfamiliar | 4.2<br>3.8 | 0.259 | **0.014**** | —— | **0.655** |
| Q2 | Familiar<br>Unfamiliar | 4.633<br>4.467 | 0.182 | 0.201 | —— | 0.334 |
| Q3 | Familiar<br>Unfamiliar | 4.2<br>4.333 | 0.159 | 0.472 | —— | 0.187 |
| Q4 | Familiar<br>Unfamiliar | 4.333<br>4.2 | **0.001***** | —— | 0.351 | 0.243 |
| Q5 | Familiar<br>Unfamiliar | 4.4<br>4.367 | 0.866 | 0.835 | —— | 0.054 |
| Q6 | Familiar<br>Unfamiliar | 4.267<br>4.433 | 0.898 | 0.290 | —— | 0.275 |
| Q7 | Familiar<br>Unfamiliar | 4.2<br>4.067 | **0.099*** | 0.366 | —— | 0.235 |
| Q8 | Familiar | 4.333 | 0.336 | 0.456 | —— | 0.194 |
| | Unfamiliar | 4.2 | | | | |
| Q9 | Familiar<br>Unfamiliar | 4.5<br>4.167 | 0.029** | ____ | **0.031**** | **0.572** |
| Q10 | Familiar<br>Unfamiliar | 4.433<br>4.2 | 0.149 | 0.1677 | | 0.361 |

**Note**: ***, **, and * denote significance levels of 1%, 5%, and 10%, respectively.

### 4.4.3 The influence analysis of gender (male vs female)

The statistical analysis was conducted to examine gender differences in performance across 10 questions (Q1-Q10) between two groups—Male and Female. The normality test has already been carried out in *Section* 4.4.2, and it was considered that all the data was essentially acceptable as normally distributed, so it was straightforward to start with the other tests. Homogeneity of variance test was first performed to assess whether the variances between the two groups were equal. In cases where the homogeneity assumption was met (Q1, Q2, Q4-Q9), an independent samples t-test was used to determine if there were significant differences between the groups. For Q3 and Q10, where the homogeneity assumption was violated, Welch's t-test, which does not assume equal variances, was applied. The results are shown in Table 2.

The analysis reveals that there were no significant gender differences in performance for most of the questions, as the *p*-values for the majority of questions were greater than 0.1. Notably, a significant difference was found for Q2, with a *p*-value of 0.063, which is slightly below the conventional threshold of 0.1. However, Cohen's d value for this difference was 0.491, indicating a small effect size. This suggests that while the difference was statistically significant, its practical significance is limited.

In summary, these findings indicate that gender did not have a substantial impact on performance across the majority of the questions. Although statistical significance was found for Q2, the small effect size suggests that this difference is not of considerable practical importance. Therefore, gender does not appear to be a major factor influencing the results of this study.

Table 2 The influence analysis of gender.

| Questions | Factor | Average score | Homogeneity test (*p*-value) | t-Test (*p*-value) | Welch's t-Test (*p*-value) | Cohen's d value |
|---|---|---|---|---|---|---|
| Q1 | Male<br>Female | 4.162<br>3.929 | 0.652 | 0.422 | 0.424 | 0.209 |
| Q2 | Male<br>Female | 4.438<br>4.679 | 0.088* | **0.063*** | ---- | 0.491 |
| Q3 | Male<br>Female | 4.375<br>4.143 | **0.006***** | ---- | 0.201 | 0.329 |
| Q4 | Male<br>Female | 4.281<br>4.25 | 0.658 | 0.828 | ---- | 0.057 |
| Q5 | Male | 4.406 | 0.932 | 0.760 | ---- | 0.079 |
| | Female | 4.357 | | | | |
| Q6 | Male | 4.344 | 0.310 | 0.933 | ---- | 0.022 |
| | Female | 4.357 | | | | |
| Q7 | Male | 4.188 | 0.233 | 0.433 | ---- | 0.204 |
| | Female | 4.071 | | | | |
| Q8 | Male | 4.375 | 0.466 | 0.193 | ---- | 0..341 |
| | Female | 4.143 | | | | |
| Q9 | Male | 4.375 | 0.607 | 0.571 | ---- | 0.148 |
| | Female | 4.286 | | | | |
| Q10 | Male | 4.25 | **0.016**** | ---- | 0.388 | 0.219 |
| | Female | 4.393 | | | | |

**Note**: ***, **, and * denote significance levels of 1%, 5%, and 10%, respectively.

## 5. Discussion

### 5.1 *Broader map color knowledge*

The current system demonstrates substantial capabilities in utilizing LLM for generating map color schemes based on predefined parameters. However, there remains considerable potential to expand the system's functionality by incorporating a broader and more nuanced understanding of map color knowledge. As remarked by P3 'The number of available themes now is a bit limited'. This would allow for the development of more sophisticated and contextually relevant color schemes, addressing a wider range of map types and user needs.

(1) Integration of Data-Driven Insights. The system could further expand its capabilities by integrating data-driven insights into the color generation process. By analyzing the themes of maps and the data they represent, the system could adapt its color generation to be more suitable for specific themes or data. For instance, by analyzing the underlying data, the system could apply more suitable color schemes, improving the clarity and utility of the final map.

(2) Context-Aware Color Generation. A key aspect of enhancing the system's color knowledge is the integration of context-aware color generation. While the existing system generates color schemes primarily based on standard color theory, it could benefit from considering additional factors such as cultural preferences, color blindness, and regional color conventions. For example, the use of certain color palettes may vary across different cultures and geographic regions, with specific colors having symbolic meanings (e.g., red representing danger or urgency in some regions). By incorporating region-specific and culturally relevant color data, the system could generate more culturally appropriate and effective color schemes for a global user base. Incorporating such nuanced knowledge could increase the system's versatility, ensuring that the generated maps are not only visually appealing but also relevant and culturally sensitive.

### 5.2 *Extend multimodal and user interaction*

While the system currently generates map colors based on text input and preset parameters, there is significant potential to expand its interactivity and multimodal capabilities. Future versions of the system could allow users to interact with the tool through more diverse modes, such as visual feedback and voice commands. This would provide users with a more intuitive experience, particularly for those who may find textual interfaces less comfortable or accessible. For example, many users now prefer generating color schemes by providing an image, from which colors are then transferred to their maps (Wu et al., 2022). However, current image-to-map color transfer methods are typically end-to-end processes, which lack flexibility. Integrating this feature into our system, alongside the existing text input option, could offer greater flexibility, allowing users to apply colors directly from an image to a map while receiving immediate visual feedback that can aid decision-making. Two modes were planned to be added to this system:

(1) **Visual interaction**. Unlike previous approaches where the input image is solely used as a color source, our system will allow users to interactively select specific objects or areas within the image for color extraction. This selective approach gives users more control over which elements of the image are used to generate the color scheme for the map. Additionally, to enhance personalization, users will also have the option to combine text input with an image input, allowing them to refine the color selection process by specifying textual preferences alongside visual cues. This capability leverages the power of advanced visual large language models, which can understand and process images. For example, models like CLIP can describe the content of an image, and DINO v2 can segment the image and recognize objects. This flexibility allows for the more nuanced and context-aware color generation, catering to both expert and non-expert users.

(2) **Voice interaction**. Similar to text-based interactions, voice input would be processed in two stages: the spoken input would first be converted into text through speech recognition technology, and the text would then be used to drive the system's response. Advances in large language models, such as GPT-4, now support voice-based interactions, which can be seamlessly integrated into the system. Given that voice interactions may take slightly longer to process than text inputs, careful design of the user interface is necessary to avoid frustration. To enhance responsiveness, the system could stream output progressively as it receives voice commands, providing users with immediate feedback while they interact with the tool. Furthermore, the system could offer prompt acknowledgments, such as confirming receipt of the command and displaying intermediate results as they are generated.

### 5.3 *Improve the system via open user evaluation*

Though we evaluated the usability of the proposed system with a dataset across 60 participants, the sample size and test cases are still limited. It is crucial to further test the system in a wider range of use scenarios and with a more diverse group of users to better understand its limitations and areas for improvement. The current evaluation provides valuable insights, but additional testing across different contexts—such as varying map data, user expertise levels, and geographical regions—is necessary to assess the system's generalizability and effectiveness in real-world applications. Furthermore, while the system is functional, it is still in the early stages of development, and significant opportunities exist to enhance its performance, versatility, and overall user experience. As P39 claimed 'The AI has shown a decent understanding capability, and the generated color schemes basically meet the requirements, but the functionality is still relatively rudimentary.' Continuous development is critical to ensure the tool's long-term success and relevance.

## 6. Conclusion

We have introduced a novel system for choropleth map color design, which leverages a large language model (LLM) to align color schemes with user intentions and data semantics. This system addresses the limitations of traditional mapping tools by providing a more adaptable and user-centric approach to map design. Through a three-stage process—Data Classification, Color Concept Design, and Color Scheme Design—the system integrates design principles and color psychology to produce relevant and visually effective color schemes. The interactive interface further enhances the system's usability, allowing users to customize and refine their color choices based on specific needs and preferences. Results from user studies highlight the tool's high usability, flexibility, and alignment with user expectations, with participants emphasizing its efficiency and ease of use. However, it is important to note that the system is still in its early stages, and several areas require improvement. Future work will focus on enhancing the accuracy of color mapping in diverse contexts and expanding the interaction modes to provide users with a more intuitive experience. Despite its current limitations, we believe this work provides valuable insights into how LLM can be used to enhance map design processes, and we look forward to continued improvements that will support creative and data-driven decision-making in spatial data visualization.

## Disclosure statement

No potential conflict of interest was reported by the author(s).

## Funding

This research was financially supported by the National Natural Science Foundation of China [No. 42171438].